\documentclass[twocolumn,showpacs,preprintnumbers,amsmath,amssymb]{revtex4}
\usepackage{dcolumn}
\usepackage{bm}
\usepackage{graphicx}
\def\rmsub#1#2{#1_{\mbox{\tiny #2}}}	    

\def \det{\mathop{\rm det}}
\def \tr{\mathop{\rm tr}}
\def \ln{\mathop{\rm ln}}

\def \seff{\mathop{S_{\rm eff}}}

\def \hmc{HMC}
\def \rhmc{RHMC}
\def \m{\mathcal M}

\def\dt{\delta\tau}			    
\def\dH{\delta H}			    
\def\md{MD}				    
\def\hmc{HMC}				    
\def\rhmc{RHMC}				    
\def\rmsub#1#2{#1_{\mbox{\tiny #2}}}	    
\def\opt#1{\rmsub{#1}{opt}}		    
\def\pacc{\rmsub{P}{acc}}		    

\def\k{\kappa}				    

\def\tD{\mbox{D}\kern-0.65em\raise0.15ex\hbox{/}\kern0.15em} 
\def\sD{\mbox{\scriptsize D}\kern-0.5em\raise0.15ex\hbox{\scriptsize/}}
\def\ssD{\mbox{\tiny D}\kern-0.42em\raise0.15ex\hbox{\tiny/}}
\def\Dslash{{\mathchoice{\tD}{\tD}{\sD}{\ssD}}}
\def\dslash{\hbox{\(\partial\)}\kern-0.5em\raise0.15ex\hbox{/}} 

\def\Nmv{N_{\mbox{\tiny{mv}}}}
\def\nth{n^{\mbox{\tiny{th}}}}
\def\rational#1#2{{\mathchoice{\textstyle{#1\over#2}}%
  {\scriptstyle{#1\over#2}}{\scriptscriptstyle{#1\over#2}}{#1/#2}}}
\def\half{\rational12}                      

\def\tint#1{\tau_{\mbox{\tiny{#1}}}}

\def\Sg{\rmsub{S}{g}}
\def\Sf{\rmsub{S}{f}}

\def\ratdeg{l}
\def\ratDeg#1{l_{\mbox{\tiny{#1}}}}

\begin{document}

\preprint{APS/123-QED}

\title{Accelerating Dynamical Fermion Computations using the \\ Rational Hybrid
Monte Carlo (RHMC) Algorithm\\ with Multiple Pseudofermion Fields}

\author{M. A. Clark}
  \email{mikec@bu.edu}
  \affiliation{%
    Center for Computational Sciences,  Boston University\\
    3 Cummington Street, Boston,  MA 02215,  United States of America}
\author{A. D. Kennedy}
 \email{adk@ph.ed.ac.uk}
  \affiliation{
    School of Physics, The University of Edinburgh\\
    Mayfield Road, Edinburgh, EH9 3JZ, United Kingdom}
\date{August 18, 2006}

\begin{abstract}
  \noindent There has been much recent progress in the understanding and
  reduction of the computational cost of the Hybrid Monte Carlo algorithm for
  Lattice QCD as the quark mass parameter is reduced. In this letter we present
  a new solution to this problem, where we represent the fermionic determinant
  using \(n\) pseudofermion fields, each with an \(\nth\) root kernel. We
  implement this within the framework of the Rational Hybrid Monte Carlo (RHMC)
  algorithm. We compare this algorithm with other recent methods in this area
  and find it is competitive with them.{\parfillskip=0pt\par}
\end{abstract}


\maketitle

The motivation for this work is the need for faster algorithms to perform
lattice QCD calculations near the chiral limit. To include the effects of
fermions in our calculations we are required to invert the Dirac operator,
which is a very large sparse matrix: as the fermion mass is decreased the
computational cost increases with the condition number of the matrix
\(\k(\m)\). Lattice QCD calculations involve the integration of Hamilton's
equations of motion, and the fermionic force acting on the gauge fields also
increases with decreasing fermion mass. To maintain the stability of the
integrator the integration step size must be reduced, thus increasing the cost.
In this paper we shall not address the first of these problems, but shall
introduce a method to bring the latter under control. This work is related to
similar work by Hasenbusch~\cite{Hasenbusch:2001ne}, but our method requires
less parameter tuning.

When performing a lattice QCD simulation, we desire gauge field configurations
\(U\) distributed according to the probability density
\begin{equation*}
  P(U) = \frac{1}{Z} e^{-\Sg(U)} \det\m(U),
\end{equation*}
where \(\Sg\) is the pure gauge action and \(\det\m\) is the determinant of the
Dirac operator \(\m=(\Dslash+m)^\dagger(\Dslash+m)\), which appears after
integrating out the Grassmann-valued quark fields. The operator \(\Dslash\) is
the discretized covariant derivative, and \(m\) is the fermion mass. We
represent the determinant as a pseudofermion Gaussian functional integral
(\(\det\m \propto \int d\phi\,d\phi^\dagger\, \exp{\left(-\phi^\dagger\m^{-1}
\phi\right)}\)), giving the probability density
\begin{equation*}
  P(U,\phi) = \frac{1}{Z'} e^{-\Sg(U) - \Sf(\m)} = \frac{1}{Z'}e^{-\seff}.
\end{equation*}

Almost all techniques for generating gauge configurations consist of variants
of Hybrid algorithms, these being algorithms which combine momentum and
pseudofermion refreshment heatbaths with molecular dynamics (\md) integration
of the gauge field. The latter is done through the introduction of a
``fictitious'' momentum field \(\pi\), with which we define the Hamiltonian \(H
= \half\pi^2 + \seff\). The gauge fields can then be allowed to evolve for a
time \(\tau\) by integrating Hamilton's equations. Hybrid algorithms are
ergodic and their fixed point is close to but not precisely the desired one.
This discrepancy is due to the inexact integration of Hamilton's equations: the
Hamiltonian is conserved to \(O(\dt^k)\), where \(k\) is determined by the
order of the integration scheme used. Hybrid algorithms which use area
preserving and reversible (symplectic and symmetric) integrators can be made
exact through the addition of a Metropolis acceptance test at the end of the
\md\ trajectory, which stochastically corrects for these errors. The Hybrid
Monte Carlo algorithm (\hmc) algorithm~\cite{Duane:1987de} is the {\it de
facto} method for generating the required probability distribution, of which a
single update consists of the following Markov steps
\begin{itemize}
\item Momentum refreshment heatbath using Gaussian noise
  (\(P(\pi)\propto e^{-\pi^2/2}\)).
\item Pseudofermion refreshment (\(P(\phi)\propto (\Dslash+m)^\dagger\eta\),
  where \(P(\eta)\propto e^{-\eta^\dagger \eta}\)).
\item MDMC, which consists of
  \begin{itemize}
  \item \md\ trajectory consisting of \(\tau/\dt\) steps.
  \item Metropolis accept/reject with probability \(\pacc = \min(1,e^{-\dH})\).
  \end{itemize}
\end{itemize}
When updating the momentum there are contributions to the force from both the
pure gauge part of the action \(\partial \Sg / \partial U\), and the fermionic
part \( \partial \Sf / \partial U\). As the fermion mass is decreased the
latter becomes the dominant contribution. To avoid an instability in the
integrator and to maintain a non-negligible acceptance rate we must reduce the
step size \(\dt\): this makes the computation very expensive since at every
{\md} step we must solve a large system of linear equations.

Since the fermion determinant is represented using a single pseudofermion
configuration selected from a Gaussian heatbath, the variance of this
stochastic estimate will lead to statistical fluctuations in the fermionic
force: the pseudofermionic force may be larger than the exact fermionic force,
which is the derivative \(\partial\tr\ln\m(U)/\partial U\) with respect to the
gauge field \(U\). This means that the pseudofermionic force may trigger the
instability in the integrator even though the exact force would not.

An obvious way of ameliorating this effect is to use \(n>\nobreak1\) 
pseudofermion fields to sample the functional integral: this is achieved simply
by writing
\begin{eqnarray*}
  \det\m & = &[\det\m^{1/n}]^n \\
  & \propto & \prod_{j=1}^n d\phi_j\, d\phi^\dagger_j\,
    \exp{\left(-\phi^\dagger_j\m^{-1/n}\phi_j \right)},
\end{eqnarray*}
that is introducing \(n\) pseudofermion fields \(\phi_j\) each with kernel
\(\m^{-1/n}\).

It is well known that the instability in the integrator is triggered by
isolated small modes of the fermion kernel~\cite{Joo:2000dh}. These modes are
of magnitude \(O(1/m^2)\) with the standard kernel; with our multiple
pseudofermion approach these modes are now \(O(1/m^2)^{1/n}\), and so are
vastly reduced in magnitude. We would thus expect that the instability is
shifted to occur at a far greater integrating step size.

If we make the simple-minded estimate that the magnitude of the fermion force
\footnote{Strictly speaking this is an ``impulse'' rather than a force.} is
proportional to the condition number of the fermion matrix multiplied by the
step size, then we can find the optimum value of \(n\). We must keep the
maximum force fixed so as to avoid the instability in the integrator, so we may
increase the integration step size to \(\dt'\) such that \(n\k(\m)^{1/n}\dt' =
\k(\m)\,\dt\), where we have used the fact that \(\kappa(\m^{1/n}) =
[\kappa(\m)]^{1/n}\). At constant trajectory length and acceptance rate, and
hence constant autocorrelation times, the cost of an \hmc\ trajectory is
proportional to the ratio \(n/\dt'\), and thus is minimized by choosing \(n\)
so as to minimize \(n\,\dt/\dt' = n^2\k(\m)^{\frac1n-1}\), which leads to the
condition \(\opt{n} = \half\ln\k(\m)\), corresponding to cost reduction by a
factor of \(n\,\dt/\dt' = \left[e\ln\k(\m) \right]^2/\left[4\k(\m)\right]\).

Our method is to apply the Rational Hybrid Monte Carlo (\rhmc)
algorithm~\cite{Clark:2003na} to generate gauge field and pseudofermion
configurations distributed according to the probability density
\begin{equation*}
  P(U,\phi_1,\ldots,\phi_n) = \frac1Z \exp{\left[-\Sg(U) - \Sf(\m,n)\right]}
\end{equation*}
where \(\Sf(\m,n) = \sum_{j=1}^n \phi^\dagger_j \m^{-1/n}\phi_j\). Optimal
Chebyshev rational approximations are used to evaluate the matrix functions,
and we proceed as we would for conventional \hmc~\cite{Duane:1987de}. If
written in partial fraction form \(r(\m) = \sum_{k=1}^\ratdeg \alpha_k/(\m+
\beta_k)\), the rational function can be evaluated using a multi-shift
solver~\cite{Frommer:1995ik,Jegerlehner:1996pm}. The resulting computational
cost per pseudofermion field very similar to \hmc\ \footnote{There is a small
additional overhead from having to perform a matrix inversion to evaluate each
heatbath (where we have to calculate \(\m^{1/2n}\eta\)) once per trajectory.}
since the shifts \(\beta_k\) are all positive, the most costly shift being that
which is closest to zero. Remarkably, all the coefficients \(\alpha_k\) are
also positive, so the procedure is numerically stable.

At this point it is worth comparing our method to the multiple pseudofermions
through mass preconditioning method, or the so called Hasenbusch
trick~\cite{Hasenbusch:2001ne}. In the latter, the fermion determinant is
written \(\det \m = \det \hat{\m} \det[\m/\hat{\m}]\), where the mass parameter
used in \(\hat{\m}\) is larger than that in \(\m\).  The original idea behind
this method was to tune the mass of the dummy operator \(\hat{\m}\) such that
the operators had a similar condition number~\cite{Hasenbusch:2002ai}. An
increase in step size would then be possible for the same reasons given
above. The advantage with this method compared to \rhmc\ is that the extra
operators introduced are heavier, and hence cheaper to evaluate compared to the
original kernel. Recently, larger speedups have been found through tuning the
dummy mass(es) such that the action constituents with the greatest forces are
those which are cheapest to evaluate, i.e., the heaviest~\cite{Urbach:2005ji}.
A multi-level integration scheme~\cite{Sexton:1992nu} is then used which
evaluates the cheaper and dominant forces more frequently than the more
expensive and smaller forces. Tuning the mass parameters with both of these
methods requires some effort, and even more so as further dummy operators are
introduced. This compares to our \rhmc\ method which requires no tuning of the
extra operators since all operators are identical.

\rhmc\ has the added virtue in that it allows the inclusion of less fermions
than are described by the kernel \(\m\) (typically this represents two
fermions). For example to simulate full QCD, we are required to include the
contribution from the light quark pair (at present we always assume
\(m_u=m_d\)) and the strange quark. Traditionally the inclusion of the strange
quark has proved problematic, and the use of an inexact
algorithm~\cite{Gottlieb:1987mq} has been required. An alternative has been to
use polynomial approximations, but such an approach requires either a very
large degree polynomial (\(>O(1000)\) for light quarks), or a correction step
which is applied with the acceptance test or when making
measurements~\cite{deForcrand:1996ck,Frezzotti:1997ym}. \rhmc\ allows the
strange quark to be included simply through the use of the rational approximation
\(\sqrt\m \approx r(\m)\), and because of the high accuracy of rational
approximations, no correction step is required.

\begin{figure}
  \scalebox{0.3}[0.3]{\includegraphics{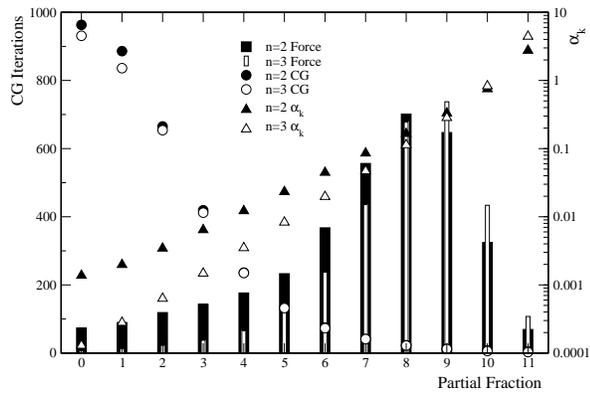}}
  \caption{\label{fig:force} The variation of the force magnitude (\(L_2\)
    norm), conjugate gradient (CG) cost and \(\alpha_k\) parameter for each
    partial fraction with \(n=2,3\) pseudofermions (\(24^3\times32\) lattice,
    \(\beta=5.6\), \(\kappa=0.15800\), CG residual \(=10^{-6}\) for all
    shifts).}
\end{figure}

An important observation that has been made with prior use of \rhmc\ is that
the derivatives of the partial fractions have vastly different
magnitudes~\cite{Clark:2005sq}. In Figure \ref{fig:force} we show the variation
of magnitude of the force with each partial fraction, in order of increasing
shift. The surprise is that the smallest shifts contribute least to the total
fermion force. Also included on the plot is the number of conjugate iterations
required in the multi-shift solver to reduce the residual for each shifted
system by six orders of magnitude relative to the source. The most expensive
constituents of the fermion force actually have the smallest magnitude. This
effect may in part due to the fact that the density of states of the Dirac
operator is greatest in the bulk, but principally because of the nature of the
rational coefficients. This latter effect is enhanced as \(n\) is increased
because the coefficients \(\alpha_k\) become smaller for light shifts, and
larger for heavier shifts (see Figure \ref{fig:force}).

We can make use of this observation by two methods. In the spirit
of~\cite{Urbach:2005ji} we can construct a multi-timescale numerical integrator
that assigns a larger step size to the small shifts, the ratio of the two
step sizes being chosen such that the product \(F\dt\) is the same. In practice
we have found the simpler approach where the smaller shifts are given a looser
stopping condition than the heavier shifts while keeping the step size the same
for all shifts, to be a more effective approach. It is important to mention
that loosening the stopping condition of the poles has no effect on the
reversibility of the molecular dynamics~\cite{Joo:2000dh}. Here we are
loosening the stopping condition of the smaller shifts which are less important
for evaluating the total fermion force.

\begin{figure}
  \scalebox{0.25}[0.25]{\includegraphics{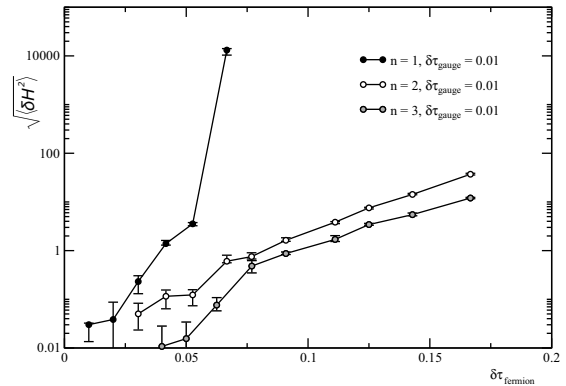}}
  \caption{\label{fig:dH} The variation of \(\dH\) with step size for
    \(n=1,2,3\), the integrating step size has been fixed at approximately
    \(\dt=0.01\) (Wilson gauge action, Staggered fermions, \(16^4\) lattice,
    \(\beta=5.76\), \(m=0.005\)).}
\end{figure}

To test our hypothesis that the use of multiple pseudofermions removes the
integrator instability, we produced the data shown in Figure \ref{fig:dH} using
a relatively modest lattice volume. On a logarithmic scale we plot the value of
\(\langle \dH^2 \rangle ^{1/2}\) versus the step size for \(n=1,2\) and~\(3\)
pseudofermions. We used a multi-timescale integrator to isolate the effect of
the fermions from that of the gauge action.  For \(n=1\) when the step size
reaches \(\dt=0.066\), \(\dH\) explodes by four orders of magnitude, this
corresponds to the value for which the instability is triggered. For \(n=2\)
and~\(3\) not only is energy conservation better, but also the instability has
been removed.

In conventional \hmc\ the second order leapfrog integrator is usually found to
be optimal, and the use of higher order integrators are found to decrease
performance. Higher order integrators are more susceptible to the instability
discussed before because they are constructed from longer sub-steps than the
original leapfrog step. If multiple pseudofermions bring the instability under
control, higher order integrators should now prove advantageous; this is indeed
found to be the case. We have tried a variety of fourth order integrators: the
fourth order Campostrini~\cite{Campostrini:1989ac} integrator and fourth order
minimum norm integrators~\cite{Takaishi:2005tz}. While these all help we have
generally found the fourth order minimum norm integrator (4MN5FV) to be
optimal. Initial investigation has not found any gain from using a sixth order
integrator, though this (or an even higher order integrator) should not be
ruled out from future investigation.

\begin{table}[htb]
  \advance\tabcolsep by2pt
  \begin{tabular}{lrrrcll} \hline\hline\noalign{\vskip1pt}
    \hfil\(\kappa\)\hfil & \(n\) & \(\ratDeg{md}\) & \(\ratDeg{mc}\)
    & Integrator & \hfil\(\tau\)\hfil & \hfil\(\dt\)\hfil \\[1pt] 
    \hline\noalign{\vskip1pt}
    0.15750 & 2 & 11 & 15 & 2MN & 1.0 & 0.1 \\[1pt]
    0.15800 & 3 & 12 & 16 & 4MN5FV & 2.0 & 0.25 \\[1pt]
    0.15825 & 3 & 12 & 16 & 4MN5FV & 2.0 & 0.25 \\ \hline\hline
  \end{tabular}
  \caption{\label{table:parameters}Table of parameters used for this study.
    The 2MN integrator is the second order minimum norm
    integrator~\cite{Takaishi:2005tz}.  For the \md\ the smallest two shifts
    had a CG residual of \(10^{-4}\), \(10^{-5}\) for the next two, and
    \(10^{-6}\) for the remaining shifts. For all shifts the CG residuals were
    set to \(10^{-10}\) for the heatbath and action evaluations.}
\end{table}

\begin{table}[htb]
  \advance\tabcolsep by1pt
  \begin{tabular}{lllrrrr} \hline\hline\noalign{\vskip2pt}
    &&&& \multicolumn{3}{c}{\(\tint{plaq}\cdot\Nmv\times10^{-4}\)} \\[2pt]
    &&& \multicolumn{1}{c}{\(\Nmv\times\)}
    & \multicolumn{1}{c}{\raise-2pt\hbox{This}}
    & \multicolumn{1}{c}{Ref.} & \multicolumn{1}{c}{Ref.} \\[1pt]
    \multicolumn{1}{c}{\(\kappa\)} & \multicolumn{1}{c}{\(\langle A\rangle\)}
    & \(\tint{plaq}\) & \multicolumn{1}{c}{\(10^{-4}\)} 
    & \multicolumn{1}{c}{\raise1pt\hbox{paper}}
    & \multicolumn{1}{c}{\cite{Urbach:2005ji}}
    & \multicolumn{1}{c}{\cite{Orth:2005kq}} \\
    \hline\noalign{\vskip1pt}
    0.15750 & 0.755(9)  & 7(1)   & 1.37  & 9.59 & 9.00 & 19.075 \\[1pt]
    0.15800 & 0.935(10) & 4.9(8) & 3.9  & 19.11 & 17.36 & 128.000 \\[1pt]
    0.15825 & 0.911(12) & 4.7(6) & 11.2 & 52.50 & 56.50 
    & \multicolumn{1}{c}{---} \\
    \hline\hline
  \end{tabular}
  \caption{\label{tabl:results}Table of results comparing \rhmc\ with
    conventional \hmc\ and the multiple timescale mass preconditioning
    presented in~\cite{Urbach:2005ji}. Our measure of cost is the product of
    the integrated autocorrelation of the plaquette
    \(\tau_{\mbox{\tiny{plaq}}}\) with the number of Dirac operator
    applications per trajectory \(\Nmv\). }
\end{table}

To test the efficiency of our final algorithm we choose the same parameters
that have been used in recent publications~\cite{Luscher:2005rx,
Urbach:2005ji}: namely a \(24^3\times32\) lattice, using a Wilson gauge action
(\(\beta=5.6\)) together with unimproved Wilson fermions with three different
mass parameters.

A summary of our results compared to multi-timescale mass
preconditioning~\cite{Urbach:2005ji} and conventional \hmc~\cite{Orth:2005kq}
can be seen in Table \ref{tabl:results}. Our algorithm is very similar in
performance to that presented in~\cite{Urbach:2005ji}, and both of these
algorithms are clearly superior to conventional \hmc\ as expected. The results
are also comparable with those of L\"uscher~\cite{Luscher:2005rx}, but are far
easier and more efficient to implement especially on fine grained parallel
computers.

\section*{Conclusions}

In this letter we have presented a simple improvement to the \hmc\ algorithm to
reduce the computation required for Lattice QCD calculations. This method leads
to a large gain in performance relative to the conventional algorithm.  At the
physical parameters analyzed, our method is competitive with the mass
preconditioning method presented in~\cite{Urbach:2005ji}; moreover, the use of
the RHMC algorithm permits the easy introduction of single quark flavors. In
practice it is often advantageous to combine mass preconditioning with the
present \(\nth\)~root method. The benefit that is gained from the improved HMC
algorithms increases as the quark mass is reduced, and in this regime it would
be interesting to further compare these algorithms.  As the lattice volume is
increased, we expect our method to prove more advantageous because of the
improved volume dependence of higher order integrators (with a second order
integrator, the cost of HMC is expected to scale \(V^{5/4}\), whereas with a
fourth order integrator the cost is expected to scale
\(V^{9/8}\)~\cite{Creutz:1989wt}).

The importance of these results is that the cost to generate gauge field
configurations with light fermions, which is the most costly part of lattice
field theory computations, has been drastically reduced.  This corresponds to
more than a four fold decrease in computer time.  These techniques also promise
to lead to similar improvements in other fields where pseudofermion techniques
are used for fermionic Monte Carlo computations.

\section*{Acknowledgements}
We wish to thank Carsten Urbach for providing thermalized lattices for this
work.

The development and computer equipment used in this calculation were funded by the 
U.S. DOE grant DE-FG02-92ER40699, PPARC JIF grant PPA/J/S/1998/00756 and by 
RIKEN. This work was supported by PPARC grants PPA/G/O/2002/00465, PPA/G/S/- 
2002/00467 and PP/D000211/1.


\end{document}